\documentclass[8.5pt,twoside,twocolumn]{article}
\usepackage[super,sort&compress,comma]{natbib} 
\usepackage{mhchem}
\usepackage{times,mathptmx}
\usepackage{sectsty}
\usepackage{balance} 

\usepackage{graphicx} %eps figures can be used instead
\usepackage{lastpage}
\usepackage[format=plain,justification=raggedright,singlelinecheck=false,font=small,labelfont=bf,labelsep=space]{caption} 
\usepackage{fancyhdr}
\newcommand{\dif}{\mathrm{d}}

\begin{document}

\thispagestyle{plain}
\fancypagestyle{plain}{
%\fancyhead[L]{\includegraphics[height=8pt]{headers/LH}}
%\fancyhead[C]{\hspace{-1cm}\includegraphics[height=20pt]{headers/CH}}
%\fancyhead[R]{\includegraphics[height=10pt]{headers/RH}\vspace{-0.2cm}}
\renewcommand{\headrulewidth}{1pt}}
\renewcommand{\thefootnote}{\fnsymbol{footnote}}
\renewcommand\footnoterule{\vspace*{1pt}% 
\hrule width 3.4in height 0.4pt \vspace*{5pt}} 
\setcounter{secnumdepth}{5}

\makeatletter 
\def\subsubsection{\@startsection{subsubsection}{3}{10pt}{-1.25ex plus -1ex minus -.1ex}{0ex plus 0ex}{\normalsize\bf}} 
\def\paragraph{\@startsection{paragraph}{4}{10pt}{-1.25ex plus -1ex minus -.1ex}{0ex plus 0ex}{\normalsize\textit}} 
\renewcommand\@biblabel[1]{#1}            
\renewcommand\@makefntext[1]% 
{\noindent\makebox[0pt][r]{\@thefnmark\,}#1}
\makeatother 
\renewcommand{\figurename}{\small{Fig.}~}
\sectionfont{\large}
\subsectionfont{\normalsize} 

\fancyfoot{}
%\fancyfoot[LO,RE]{\vspace{-7pt}\includegraphics[height=9pt]{headers/LF}}
%\fancyfoot[CO]{\vspace{-7.2pt}\hspace{12.2cm}\includegraphics{headers/RF}}
%\fancyfoot[CE]{\vspace{-7.5pt}\hspace{-13.5cm}\includegraphics{headers/RF}}
\fancyfoot[RO]{\footnotesize{\sffamily{1--\pageref{LastPage} ~\textbar  \hspace{2pt}\thepage}}}
\fancyfoot[LE]{\footnotesize{\sffamily{\thepage~\textbar\hspace{3.45cm} 1--\pageref{LastPage}}}}
\fancyhead{}
\renewcommand{\headrulewidth}{1pt} 
\renewcommand{\footrulewidth}{1pt}
\setlength{\arrayrulewidth}{1pt}
\setlength{\columnsep}{6.5mm}
\setlength\bibsep{1pt}

\twocolumn[
  \begin{@twocolumnfalse}
\noindent\LARGE{\textbf{Case study of Rb$^+$(aq), quasi-chemical theory of  ion hydration,
  and the \emph{no split occupancies} rule}}
\vspace{0.6cm}

%\noindent\large{\textbf{Full Name,$^{\ast}$\textit{$^{a}$} Full Name,\textit{$^{b\ddag}$} and
%Full Name\textit{$^{a}$}}}\vspace{0.5cm}
%Please note that \ast indicates the corresponding author(s) but no footnote text is required. 
\noindent\large{\textbf{D. Sabo,$^{a,b}$ D. Jiao,$^{c}$ S. Varma,$^{a,d}$ L. R. Pratt,$^{e}$ and S. B. Rempe$^{a,\ast}$}}\vspace{0.5cm}

%add addresses, current address, and email
\footnotetext{\textit{$^{b}$~Department of Chemistry, New York University, New York, NY USA; E-mail: dubravko.sabo@nyu.edu}}
\footnotetext{\textit{$^{c}$~Texas Advanced Computing Center, University of Texas, Houston, TX USA; E-mail: jiao@tacc.utexas.edu}}
\footnotetext{\textit{$^{d}$~ Department of Cell Biology, Microbiology and Molecular Biology, University of South Florida, Tampa, FL USA; E-mail: svarma@usf.edu}}
\footnotetext{\textit{$^{e}$~Department of Chemical \& Biomolecular Engineering, Tulane University, New Orleans LA 70118, USA; E-mail: lpratt@tulane.edu}}
\footnotetext{\textit{$^{a}$~Center for Biological and Material Sciences, Sandia National Laboratories, Albuquerque, NM 87123, USA }}

%additional addresses can be cited as above using the lower-case letters, c, d, e... If all authors are from the same address, no letter is required

\noindent\textit{\small{\textbf{Received Xth XXXXXXXXXX 20XX, Accepted Xth XXXXXXXXX 20XX\newline
First published on the web Xth XXXXXXXXXX 200X}}}

\noindent \textbf{\small{DOI: 10.1039/b000000x}}
\vspace{0.6cm}
%Please do not change this text.

\noindent \normalsize{Quasi-chemical theory applied to ion hydration
combines  statistical mechanical theory, electronic structure
calculations, and molecular simulation, disciplines which are
individually subjects for specialized professional attention. Because it
combines activities which are themselves non-trivial, quasi-chemical
theory is typically viewed with surprise.  Nevertheless, it provides a
fully-considered framework for analysis of ion hydration.  Furthermore,
the initial calculations are indeed simple, successful, and provide new
information to long-standing experimental activities such as neutron
diffraction by hydrated ions.   Here we review quasi-chemical theory in
the context of a challenging application, Rb$^+$(aq).}
\vspace{0.5cm}
\end{@twocolumnfalse}
  ]

\section{Introduction}

%add addresses, current address, and email
\footnotetext{\textit{$^{a}$~Center for Biological and Material Sciences, Sandia National Laboratories, Albuquerque, NM 87123, USA }}
\footnotetext{\textit{$^{b}$~Department of Chemistry, New York University, New York, NY USA; E-mail: dubravko.sabo@nyu.edu}}
\footnotetext{\textit{$^{c}$~Texas Advanced Computing Center, University of Texas, Houston, TX USA; E-mail: jiao@tacc.utexas.edu}}
\footnotetext{\textit{$^{d}$~ Department of Cell Biology, Microbiology and Molecular Biology, University of South Florida, Tampa, FL USA; E-mail: svarma@usf.edu}}
\footnotetext{\textit{$^{e}$~Department of Chemical \& Biomolecular Engineering, Tulane University, New Orleans LA 70118, USA; E-mail: lpratt@tulane.edu}}
\footnotetext{\textit{$^{*}$~Correspondence may be sent to: slrempe@sandia.gov}}

Water is a chemically active liquid.   Probably the most primitive
aspect of that chemical activity is dissolution of
electrolytes, and the chemical processes based on availability of
dissolved ions.  One example is salt as a trigger of autoimmune
disease.\cite{OShea:2013,Kleinewietfeld:2013,Wu:2013,Yosef:2013}
Another example, presumably related to the first,\cite{cahalan:09,beeton:06} is the selective transport of dissolved ions
across membranes.\cite{hille,olaf:jgp,varma:jgp}  Rubidium (Rb$^+$)  is interesting
in this respect because it serves as an analog of potassium (K$^+$) that conducts
current through potassium ion channels, even though Rb$^+$ is slightly
larger (by ~0.2~\AA).\cite{mackinnon:energy:01,Tao:2009}

\begin{figure}[h]
\includegraphics[clip=true,width=3.0in]{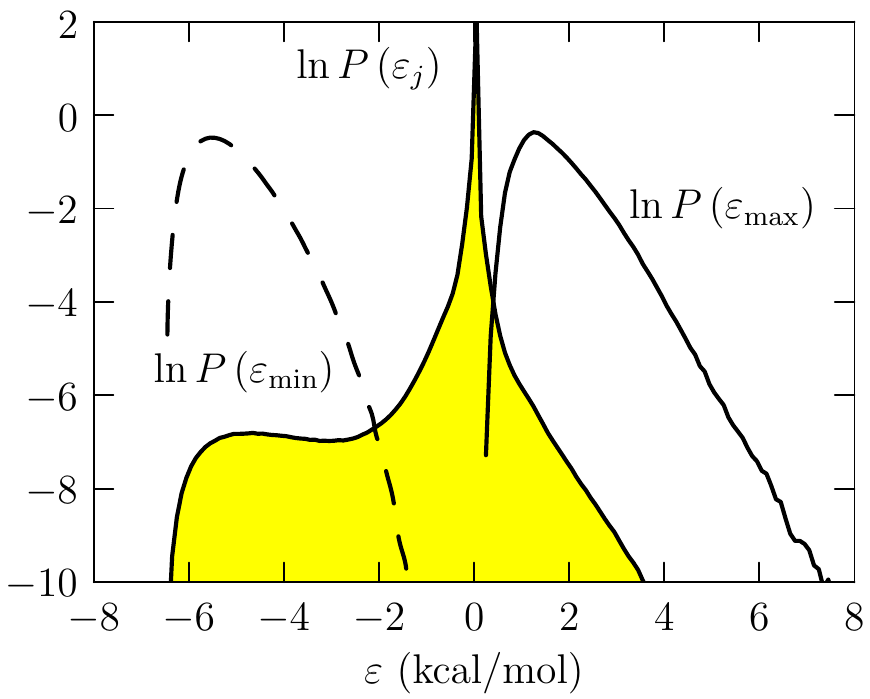}
\caption{Probability densities for pair contributions to the binding
energy of a water molecule to liquid water.\cite{Chempath:2010dq}  
Rightmost curve:  the most positive (unfavorable) pair contribution;
leftmost curve:  the most negative (favorable) contribution. The middle
curve (with yellow-shaded area) is the probability density of all the
pair interaction contributions.\cite{Stillinger:1980ws}
\label{fig:pair_eps}}
\end{figure}

In addition to this chemical activity, water has long been a serious
challenge for statistical mechanical theory of liquids, which itself is
properly almost entirely classical mechanical theory.\cite{BPP} 
The challenge presented by liquid water  is the
variety of intermolecular interactions that must be considered with a
wide range of interaction strengths
(Fig.~\ref{fig:pair_eps}).\cite{shah:144508} Those interactions include
excluded-volume repulsions, essential since liquids and liquid water are
dense materials.   Those interactions also include H-bonding
interactions that  are attractive on balance, much stronger than thermal
energies, and essential for the characteristic behavior of liquid water.

Attractive interactions that involve \emph{many} neighbors are good
candidates for treatment by mean-field
approximations.\cite{Widom:1967tz,Chandler:1983vr} Of course, if those attractive
interactions are individually weak on a thermal energy scale, that
characteristic is favorable for simple theories also.

\begin{figure}[h]
\includegraphics[clip=true,width=3.0in]{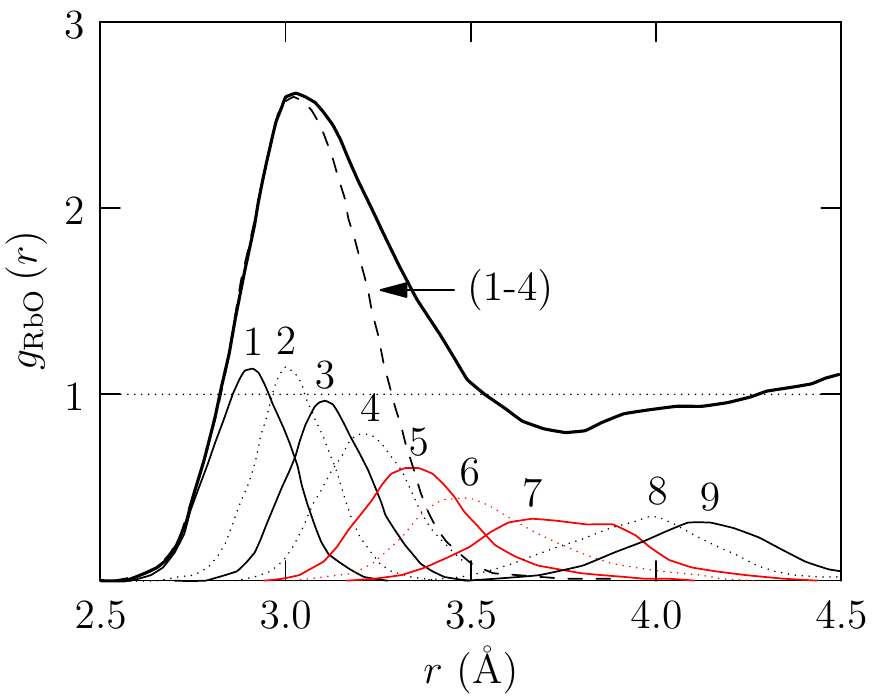}
\caption{\emph{Ab initio} molecular dynamics (AIMD) result for the
radial distribution of water oxygens about Rb$^+$, $g_{\mathrm{RbO}}$. We used VASP-version
4.2,\cite{KRESSE93A,KRESSE96A} with the Perdew-Wang (PW91) exchange-correlation
functional,\cite{PERDEW91A,PERDEW92A} a plane-wave basis set cutoff at
$36.75$~Ry with the interaction between valence and core electrons
described  by the projector augmented-wave method (PAW).\cite{BLOCHL94A}
Corrections to capture long-ranged interactions like dispersion have not been included,
and recent studies of ions in water suggest that they are not helpful.\cite{Bankura:2013}
The system consisted of one Rb$^+$ ion and $64$ waters in a
cubic simulation cell with edge length of $12.417$~{\AA} and full
periodicity. Charge balance was achieved with a neutralizing background.
All hydrogen atoms in the system were replaced by deuterium. 
The simulation was carried out for $49.28$~ps with a time step of
$0.5$~fs. The average temperature during the simulation was $347.9 \pm
4.5$~K, set intentionally high to avoid the over-structuring of pure water
observed in room temperature AIMD simulations.\cite{Schwegler:04,Vandevondele:05,Leung:PCCP06,Rempe:PCCP08}
The numbered curves show a neighborship analysis from decomposing $g_{\mathrm{RbO}}$ -- curves in red highlight waters that occupy an intermediate region distinct from the first $n$=1-4 waters
that fill in the peak of the first maximum, and more distant waters that mostly occupy the second maximum.  Results over longer length-scales are shown in
Fig.~\ref{fig:gnr} \label{fig:pgr}.}
\end{figure}

Here we consider rubidium dissolved in aqueous solution, Rb$^+$(aq), to provide a contrast with 
theories of liquids more broadly,\cite{Widom:1967tz,Chandler:1983vr} and to pursue a specific
discussion of what is yet required theoretically to treat solvated electrolytes.  The hydration free
energy of Rb$^+$(aq) is known to be roughly $-100 k_{\mathrm{B}}T$ under
standard conditions, favorable enough to dissolve simple Rb$^+$ salts, and
indeed large on the thermal energy scale. Additionally
(Fig.~\ref{fig:pgr}), the number of near-neighbors  is modest, between
four and seven.  As another contrast, the isoelectronic Kr(aq)  has
about eighteen (18) near-neighbors.\cite{Bowron:1998vl,Ashbaugh:2003bl} 
 From this comparison it is clear that the features of strong attractive
interactions and a reduced number of near-neighbors are correlated:  the
reduced number is a consequence of the crowding of near-neighbors
drawn close by the interesting attractive interactions.

This discussion suggests that we seek a way forward by focusing on the
\emph{small} number of near-neighbors, deploying direct quantum mechanical computation for
Rb(H$_2$O)$_n{}^+$ with a small number $n$ of neighbors, then stitching
those computational results into the broader theory of liquids.  That
was indeed the idea of quasi-chemical theory discussed here.   At a
formal level that theory is fully conclusive, but it is surprising that
it was not worked-out until fairly
recently.\cite{MARTINRL98A,Pratt:1998p1684,PRATT99A,PRATT02D,CPMS,rogers:ARCC} It
was also surprising that the initial applications of the theory, to
Li$^+$(aq),\cite{REMPE00A} were highly effective -- predictions of hydration free energy
matched experiment\cite{YMARCUS91A} and  {\it ab initio} molecular simulation estimates.\cite{leung:JCP09} 
Further, the initial applications provided new
information to long-standing neutron diffraction work on hydrated ions,\cite{VARMA06B} 
and indeed motivated renewed effort on those experiments.\cite{Mason:2006fd}

The initial applications of quasi-chemical theory could be understood on
a physical and intuitive basis. That encouraged the generation of simple
mimics that were less fully thought through.  A focused discussion of
the statistical thermodynamic formalities was given by Asthagiri,
\emph{et al.,}\cite{Asthagiri:2010tj} with the intention of reducing the
confusion that can result from a crowd of imitators. %impostors.
 The presentation of
Asthagiri, \emph{et al.}\cite{Asthagiri:2010tj} will be the basis of the
discussion that follows.

\section{Theory}

The primary theoretical target for quasi-chemical theory has been the
\emph{excess} (or \emph{interaction part}) of the partial molar Gibbs
free energy (or \emph{chemical potential}) of the species of interest,
here the Rb$^+$(aq).   This $\mu_{\mathrm{Rb^+}}^{\mathrm{(ex)}}$ is
indeed a basic characteristic of the solution and the Rb$^+$ ion in it,
but it is also  comparatively simple. The  potential distribution
theorem\cite{BPP} offers a partition function for evaluation utilizing
information obtained on the local environment of the ion.  It is found
that the desired free energy can be cast as
\begin{multline} 
\mu_{\mathrm{Rb^+}}^{\mathrm{(ex)}} = -kT\ln
K_{n}^{(0)}\rho_{\mathrm{H_2O}}{}^{n}  + kT \ln
p_{\mathrm{Rb^+}}\left( n\right) \\ + \mu_{\mathrm{Rb(H_2O)}_{{
n}}^+}^\mathrm{(ex)} -  {n} \mu_{\mathrm{H_2O}}^\mathrm{(ex)}~.
\label{eq:qca} 
\end{multline} 
On the right side, notice the reference to the molecular complex
$\mathrm{Rb(H_2O)}_n{}^+$ and its excess free energy
$\mu^\mathrm{(ex)}_{\mathrm{Rb(H_2O)}_n{}^+}$; the treatment of the
complex itself as a chemical constituent  is a characteristic feature of
this quasi-chemical theory.\cite{PRATT99A,BPP,CPMS,rogers:ARCC}

\begin{figure}[h]
\includegraphics[clip=true,width=3.0in]{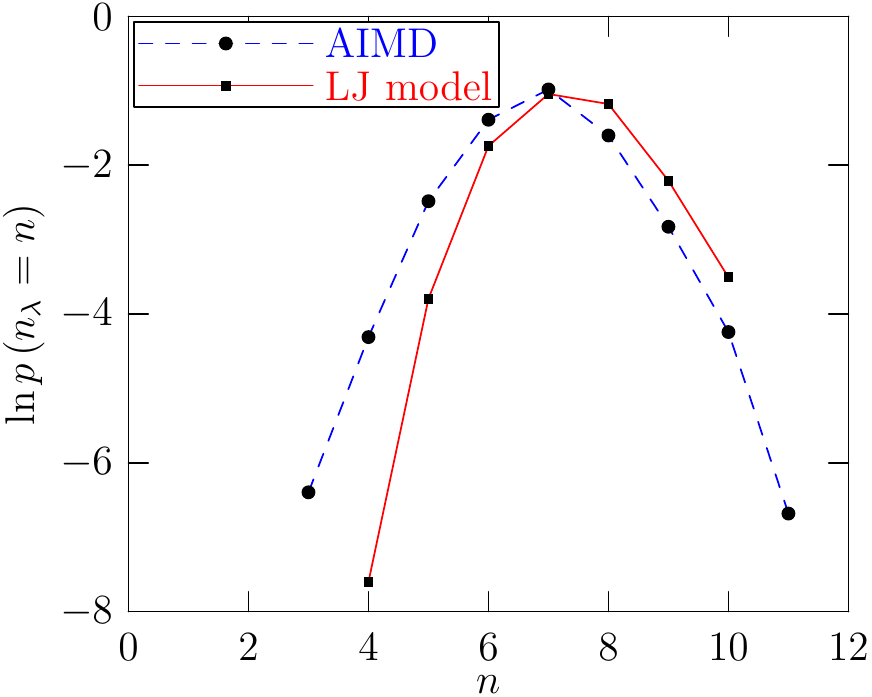}
\caption{\label{fig:lnx0_Rb} The coordination number distribution, for
the inner-shell radius of $\lambda = 3.76$~{\AA}, from the AIMD
simulations and an ``LJ model.''   For the LJ model, we used {\AA}qvist
Lennard-Jones parameters\cite{AQVIST90A} for the Rb$^+$ ion, ($\epsilon
= 1.71\times 10^{-4}$~kcal/mol, $\sigma = 5.62177$~{\AA}), and the SPC/E
potential\cite{BEREND87A} for water intermolecular interactions. We
carried out standard NVT molecular dynamics calculations using the
GROMACS\cite{GROMACS} package (version 3.1.4). The system consisted of
one Rb$^+$ and $2177$ water molecules in a ($40$~{\AA})$^3$ cell.  A
single Cl$^-$ was included for charge balance.  For
electrostatic interactions, the particle mesh Ewald technique was
implemented with Fourier spacing of $1.5$~{\AA}, a sixth-order
interpolation, a $10$~{\AA} cutoff in a direct space, and a tolerance of
$10^{-5}$. A cutoff distance of $16$~{\AA} was adopted for the
Lennard-Jones interactions.  Intramolecular geometric constraints on
water molecules were enforced by the SETTLE algorithm.\cite{KOLLMAN92A}
Data was collected during a $1.0$~ns production phase, following a
$1.0$~ns equilibration phase. The Nose-Hoover thermostat with a coupling
constant of $0.2$~ps maintained $T =
298.15$~K.\cite{NOSE84A,NOSE84B,HOOVER85A} The variation in coordination
number $n$ is substantially smaller for the LJ model data than for the
AIMD results; but remember that the AIMD results correspond to the
slightly higher $T \approx 350$~K.
} 
\end{figure}

Other features of Eq.~\eqref{eq:qca} properly fill-in a picture of
the chemical equilibrium of this complex.  The combination
$\mu_{\mathrm{Rb(H_2O)}_{{ n}}^+}^\mathrm{(ex)} -  {n}
\mu_{\mathrm{H_2O}}^\mathrm{(ex)}$ begins a free energy balance for the
association reaction
\begin{equation} \label{method2}
\mathrm{Rb}^+ + n\mathrm{H_2O}  \rightleftharpoons
\mathrm{Rb(H_2O)}_n{}^+ ~.\end{equation} 
Similarly, 
\begin{eqnarray}
K_{n}^{(0)} = \frac{\rho_{\mathrm{Rb(H_2O)}_{{
n}}{}^+}}{\rho_{\mathrm{Rb}^+}\rho_{\mathrm{H_2O}}{}^n}~,
\end{eqnarray}
with $\rho_\mathrm{X}$ the number density of species X, 
is the equilibrium ratio  for that association reaction (Eq.~\eqref{method2})
treated as in an ideal gas, \emph{i.e.,}
neglecting  interactions with the solution external to the complex; that
is signified by the superscript zero.  As emphasized
previously,\cite{BPP,PRATT99A,Asthagiri:2010tj} $K_{n}^{(0)}$ is a
well-defined few-body computational target, and the full force of
available quantum mechanical computational methods can be brought to
bear.   This has been especially relevant to treatment of transition
metal ions where d-orbital  splitting may be addressed,\cite{CPMS} and
where the simplest applications of the present theory are particularly
effective.\cite{AsthagiriD:Hydsaf,Jiao:11} Electronic charge-transfer between
ion and ligands is another effect of quantum-mechanical origin that is
transparently and simply included in these quasi-chemical approaches.
Proper accounting of electronic charge density is important for computing
absolute and relative ion binding affinities, which is useful for probing mechanisms of
selective ion binding.\cite{varma:bj10}

The remaining feature of Eq.~\eqref{eq:qca} is the probability
$p_{\mathrm{Rb^+}}\left( n\right)$ of observing $n$ ligands within a defined
inner shell (Fig.~\ref{fig:lnx0_Rb}). That inner shell is a fundamental
concept for this approach, and we return below to discuss it further. 
For now, note that if only one coordination number $n$ were ever
observed,  then $p_{\mathrm{Rb^+}}\left( n\right) = 1$ and that
contribution in Eq.~\eqref{eq:qca} would vanish.  Ultimately, that
contribution in Eq.~\eqref{eq:qca} carries the full thermodynamic
effect of whatever actual variablity of inner-shell occupancy does occur.

\begin{figure}[htbp]
\includegraphics[clip=true,width=3.2in]{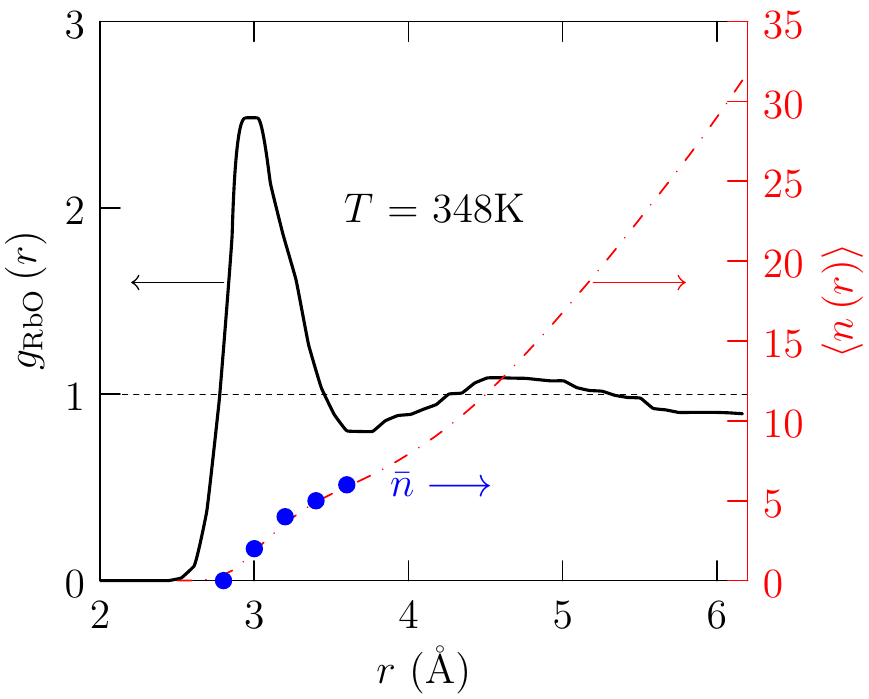}
\caption{\label{fig:gnr} Distribution $g_{\mathrm{RbO}}(r)$ of oxygen
(O) atoms radially from the Rb$^+$ ion in the AIMD simulation, emphasizing the variation of the
most probable occupancy ($\bar n$, the blue dots) of the spherical
inner shell defined by the indicated radius $r$.  $\bar n$, being
integer-valued, changes discontinuously with defining radius, but
reliably tracks the mean number of oxygen atoms within $r$, \emph{i.e.},
$\left\langle n(r)\right\rangle =$ $ 4 \pi \rho_\mathrm{O}\int_0^r
g_{\mathrm{RbO}}(x)x^2 \dif x$ for water at density $\rho_{\mathrm{0}}$.  
For example, $\left\langle n(r)\right\rangle$ = 6.86 $\pm$ 0.21 and $\bar n$ = 7 at
$r$ = 3.76~{\AA} for AIMD compared to  7.43 $\pm$ 0.22 and  7 from the LJ model.
Nevertheless, note that $\left\langle n(r)\right\rangle$ does not exhibit a convincing plateau for physical
identification of a coordination number because the principal minimum of $g_{\mathrm{RbO}}$
near $r$ = 3.76~\AA\ is weak.}
\end{figure}

Leaving the occupancy probability for general consideration, 
the left side of Eq.~\eqref{eq:qca} is independent of $n$,
and Eq.~\eqref{eq:qca} therefore describes the $n$-dependence of the probability
$p_{\mathrm{Rb^+}}\left( n\right)$. If our goal is to evaluate the free
energy, however, we can choose $n$ for our convenience.  An interesting
choice is $n = {\bar n}$, the most probable value (Fig.~\ref{fig:gnr}).
This choice makes the negative contribution $kT \ln
p_{\mathrm{Rb^+}}\left( n\right)$ as small as possible, and suggests
neglecting that contribution to obtain the convenient approximation
\begin{eqnarray} 
\mu_{\mathrm{Rb^+}}^{\mathrm{(ex)}} \approx -kT\ln
K_{{\bar n}}^{(0)}\rho_{\mathrm{H_2O}}{}^{{\bar n}} +  \mu_{\mathrm{Rb(H_2O)}_{{
{\bar n}}}^+}^\mathrm{(ex)} -  {{\bar n}} \mu_{\mathrm{H_2O}}^\mathrm{(ex)}~.
\label{eq:pqca} \end{eqnarray} 
The neglected contribution is negative, and the approximate result
Eq.~\eqref{eq:pqca} will be higher than the true free energy.  
Nevertheless, a specific evaluation of $p_{\mathrm{Rb^+}}\left( {\bar n}\right)$
can be straightforwardly extracted from standard molecular
simulations,\cite{Jiao:11CO2,Rogers:11JPCB} and thus  the significance of fluctuatations of composition
of the inner shell is obtained merely by noting the size of the
neglected $kT \ln p_{\mathrm{Rb^+}}\left( {\bar n}\right)$.  Those
results are obtained and discussed below.

\section{Application to Rb$^+$(aq)}

The primitive quasi-chemical theory, Eq.~\eqref{eq:pqca}, has indeed been
applied to several hydrated metal ions\cite{REMPE00A,REMPE01A,AsthagiriD:Abshfe,REMPE04A,svarm08,Jiao:11} 
as well as other solvation problems.\cite{AsthagiriD:HydamH,Varma:07bj,VARMA08A,SABO08A,Jiao:12}
For transition metals, as
examples, the near-neighbor water molecules are clearly located on the
basis of chemical considerations,\cite{AsthagiriD:Hydsaf} and this
theory is straightforwardly successful.  Generic procedures for those
standard cases were given by Pratt and Asthagiri.\cite{CPMS} For other
cases, some physical judgement is required, and evaluation of the
various contributions to Eq.~\eqref{eq:qca} requires analysis.   We
consider application to Rb$^+$(aq) to show how that goes.

\subsection{No split occupancies}
It is important that theories teach how to understand physical problems
in addition to reproducing numerical values of central properties. For
our present problems, that learning is focused on how to catagorize
near-neighbor water molecules to make simple theories effective.

For cases like Rb$^+$, in contrast to transition metals, the important
observations arise from the AIMD simulation results of
Fig.~\ref{fig:pgr}.   The first minimum of the radial distribution  is
remarkably mild and does \emph{not} provide a convincing identification
of an inner shell.   Instead, we consider the neighborship decomposition
of that radial distribution. We see that the 7th-most nearest neighbor ($n$=7) 
contributes to the first maximum (with negligible contribution to the peak), the second maximum, and the first
minimum.  In other words, the contribution of the 7th-most distant water
neighbor to $g(r)$ is multi-modal.

\begin{figure}
\includegraphics[clip=true,width=3.2in]{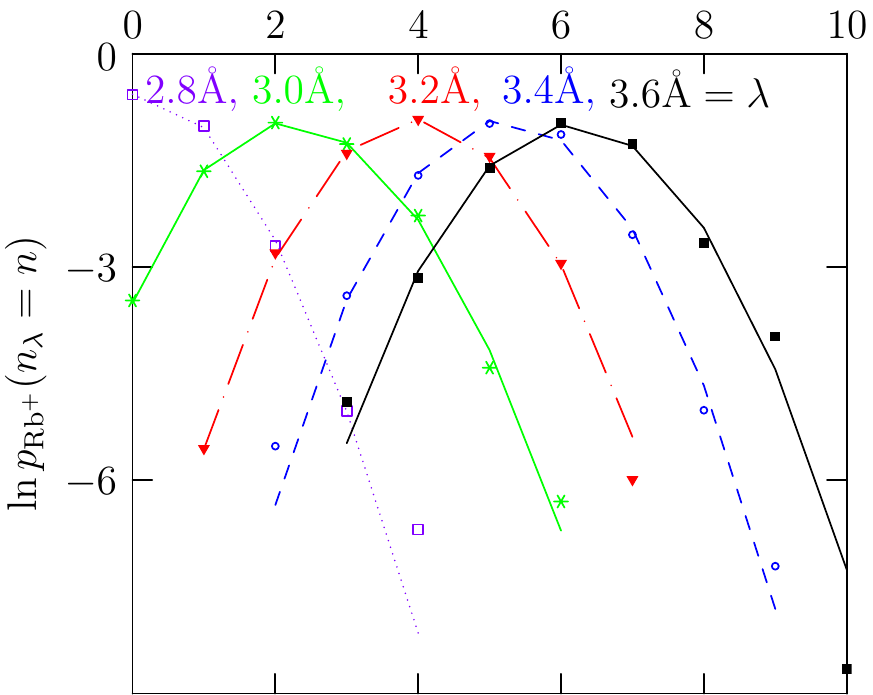}
\includegraphics[clip=true,width=3.2in]{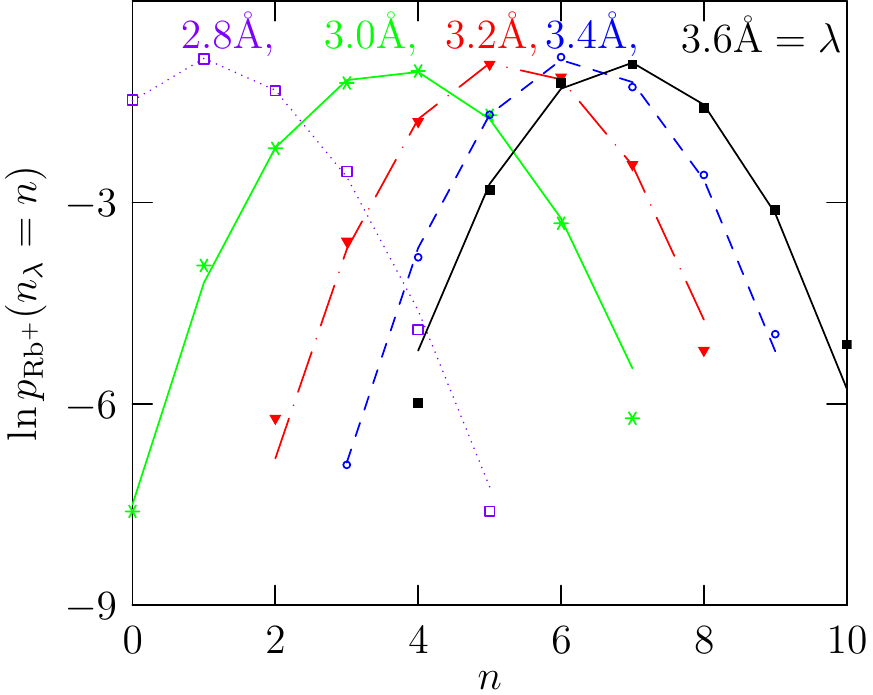}
\caption{Coordination number distributions within inner hydration shells 
defined by various $\lambda$: upper AIMD, lower LJ model.  The AIMD
results display a modest but distinct tendency 
toward lower coordination numbers.  This is possibly connected 
to the generally incorrect description of overlap repulsions by
LJ interactions.\label{fig:lnp}}
\end{figure}

To make our problem simple, we try to set an inner-shell volume so that
the 7th neighbor does not contribute.  We see from Fig.~\ref{fig:lnp}
that $p_{\mathrm{Rb^+}}\left( n=7\right)$ is particularly small for an
inner-shell radius of $\lambda = 3.2$~\AA. With the indicated choice of
inner hydration shell, we notice further that neighbors 1-4 fill-out the principal
maximum of that radial distribution function.

This lesson we will call the \emph{no split occupancies} rule.  These
neighborship analyses\cite{Mazur:1992gg} have become characteristic of
quasi-chemical theories and were used previously for Li$^+$(aq), Na$^+$(aq) and K$^+$(aq).\cite{REMPE04A,VARMA06B,Varma:07bj,svarm08,rogers:ARCC}

\subsection{Methods}

To evaluate the quasi-chemical free energy contributions to
$\mu^{\mathrm{{(ex)}}}_{\mathrm{Rb}^+}$ , we start with the first term of
Eq.~\ref{eq:pqca}.  This term gives the free energies for association of
Rb$^+$ with $n$ water ligands to form clusters within our choice of
inner-shell radius ($\lambda = 3.2$~\AA).   These clustering equilibria
take place in an ideal gas with a water density corresponding to a pressure
of 1~atm.

Gas-phase thermochemical data required for the association reactions
(Eq.~\eqref{method2}) were obtained by electronic structure calculations
using the Gaussian09 program\cite{GAUSSIAN} and density functional
theory with Becke's three-parameter exchange functional\cite{BECKE93A}
and the LYP\cite{PARR88A} electron correlation function (B3LYP). All
structures were fully optimized with a basis including polarization and
diffuse functions (6-311++G(2d,p)) on oxygen and hydrogen centers, and
the LANL2dz effective core potential and basis set on Rb$^+$.   At the
lowest-energy geometry, confirmed by zero-valued slopes of the
electronic energy with respect to atomic displacements, a standard
Hessian analysis was performed to compute normal mode vibrational
frequencies\cite{WDC,rempe:ChemEd98} using the same basis set.  Quantum
mechanical partition functions were then calculated,\cite{MCQUARRIE76} thus providing a  
determination of the free energy changes of the association reactions
due to atomic motions internal to water and the clusters at temperature $T=298$~K and
1 atm pressure. 

In a subsequent step, the \emph{cluster results} were adjusted
with a \emph{ligand replacement} contribution, $n \ln
\rho_{\mathrm{H_2O}}$, to account for the actual concentration of water ligands
in liquid water at the density $\rho_{\mathrm{H_2O}}$ =1 g/cm$^3$.  If
this density is tracked as an adjustment of the ideal gas pressure, then it
corresponds to a pressure factor of 1354~atm.

To compute the last two terms of Eq.~\eqref{eq:pqca}, $\mu_{\mathrm{Rb
(H_2O)_{n}^+}}^{\mathrm{{(ex)}}} - { n} \mu^{\mathrm{{(ex)}}}_{\mathrm{H_2O}}$,
we treated the solvent external to individual water ligands and the
inner-shell clusters as a reaction field using a polarizable continuum
model (PCM).\cite{Tomasi:05}  We subtracted the gas-phase
electronic structure energy for the $n$-coordinate cluster geometry to obtain the
desired excess free energies.  These two terms combined make up the
\emph {outer-shell electrostatics} contribution.  Finally, we evaluated 
Rb$^+$ \emph {hydration free energy}, $\mu^{\mathrm{{(ex)}}}_{\mathrm{Rb}^+}$,
 by summing the quasi-chemical components for formation of the
most probable complex (Eq.~\ref{eq:pqca}), $\mathrm{Rb(H_2O)}_{\bar n}{}^+$ .

The integral equation formalism (IEF-PCM) was implemented for the
outer-shell elecrostatics calculations.\cite{iefpcm} A radius of
3.2~{\AA} around Rb$^+$ defined the inner-shell boundary.  In addition,
default parameters were used to define hydrogen and oxygen radii used to
create the solute cavity as a set of overlapping spheres. The dielectric
constant of the outer-shell medium was set to $78.35$ to represent
liquid water.

\section{Results} \label{sec:results}

The principal maximum of the Rb-O radial distribution function
(Fig.~\ref{fig:pgr}) is near $r\approx$ 3.0~\AA.  Experimental studies report similar
results for the location of this maximum:\cite{FULTON96A,FILLIPONI03A,NEILSON00A} $2.93 \pm 0.3$~{\AA},
$2.90$~{\AA}, and $3.05$~{\AA}.

\begin{figure}
\includegraphics[clip=true,width=3.5in]{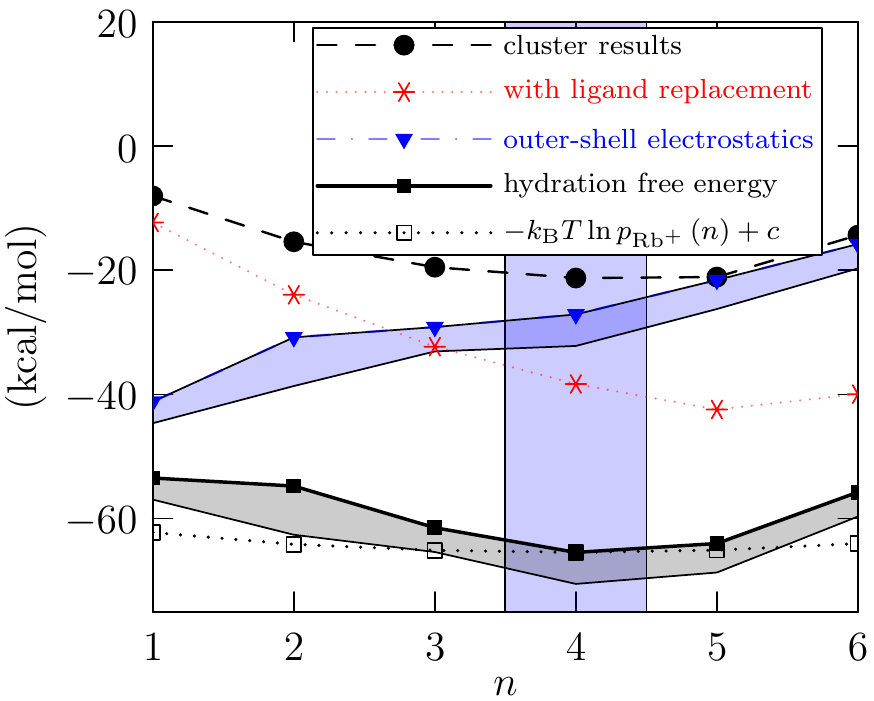}
\caption{\label{fig:qchem}
Contributions from the primitive quasi-chemical evaluation of
$\mu^{\mathrm{{(ex)}}}_{\mathrm{Rb}^+}$(aq). The predicted hydration
free energy is $-65.4$~kcal/mol when the outer-shell electrostatics
contribution is based on a \emph {single} {$n=\bar n= 4$} solvent-adapated cluster
configuration. Using instead many configurations sampled from the AIMD
simulation record, the outer-shell electrostatics contributions are
similar, but slightly lower, and predict
$\mu^{\mathrm{{(ex)}}}_{\mathrm{Rb}^+}$(aq)$=-70.5$~kcal/mol. The
modeled inner-shell occupancy distributions
$p_{\mathrm{Rb^+}}\left({n}\right)$ resemble the observations from
the AIMD simulation (Fig.~\ref{fig:lnp}, $\lambda=3.2~\AA$), plotted
with open boxes where the constant `$c$'  is adjusted to match the
single-point QCT model at $n =\bar n= 4$.   This qualitatively
satisfactory comparison achieved with both single and multiple
solvent-adapted structures (highlighted by shading) 
stands in contrast with a previous primitive
assessment\cite{Roux:2010ec} for K$^+$(aq) that is expected to be
physically similar.  That previous work did not attempt a detailed
examination of the theory, nor did it consider specifically a neighborship analysis
(Fig.~\ref{fig:pgr}), as has been long customary.\cite{REMPE04A,VARMA06B,Varma:07bj,svarm08,rogers:ARCC}}
\end{figure}

The free energy results (Fig.~\ref{fig:qchem}) show that the
coordination number $n$ = 4 is indeed the most probable within the inner hydration
shell defined by $\lambda$ = 3.2 $\AA$.  This then implies
$\mu^\mathrm{{(ex)}}_{\mathrm{Rb}^+}\mathrm{(aq)} = -65.4$~kcal/mol, which agrees
reasonably with the experimental value of
$-69.51$~kcal/mol.\cite{YMARCUS91A}

Addressing Eq.~\eqref{eq:pqca}, several approximations have accumulated.
The first of those is the neglect of population fluctuation. From
Fig.~\ref{fig:lnp}, we see that this error amounts to $kT \ln
p_{\mathrm{Rb^+}}\left( \bar{n}\right)\approx - 0.6$~kcal/mol, roughly a 1\%
error on the predicted hydration free energy. This could be easily appended to the final result, but it is not
significant here.ß

Further approximations entered to evaluate the free energies on the
right side of Eq.~\eqref{eq:pqca}.   A normal mode analyis yields harmonic frequencies that are expected to
represent the vibrational motions for small ion-water clusters.   A perturbative analysis\cite{barone:05} of
anharmonicity in the electronic energy surface confirmed that vibrations in the $\bar n = 4$ cluster are 
well-described by  normal mode analysis.  If anharmonicity were important, the corrections could easily
be included in the final result.

A serious approximation is the treatment of the external environment as
a diectric continuum when considering
$\mu_{\mathrm{Rb(H_2O)_{n}^+}}^{\mathrm{{(ex)}}} - { n}
\mu^{\mathrm{{(ex)}}}_{\mathrm{H_2O}}$. This approximation is clearly
not realistic on a molecular scale, \emph{i.e.,} the solvent is
\emph{not} actually a dielectric continuum. But in this application, the
dielectric continuum model is used for outer-shell electrostatic
effects, and thus molecular-scale inaccuracies should be less serious.  
Moveover, because of the balance of free energies in this contribution,
there is an opportunity for molecular-scale inaccuracies to cancel to
some extent.

It is also remarkable and approximate that
we  calculated $\mu_{\mathrm{Rb (H_2O)_{n}^+}}^{\mathrm{{(ex)}}}$
on the basis of a \emph {single} cluster configuration and electronic charge distribution.   Moreover, we
based the estimate of $\mu_{\mathrm{Rb (H_2O)_{n}^+}}^{\mathrm{{(ex)}}}$
on the geometry and electronic charge distribution of the cluster that has been
subtly altered by the environment.  
This inclusion of a solvent reaction field is expected \cite{rogers:ARCC} to be an improvement for simple dielectric
and Gaussian distribution theories of solvation free energies.  Further, this approach represents a
distinct change from typical  procedures used previously.\cite{rogers:ARCC}  
Although the cluster structures ($n$=1-8) are only subtly different on the basis of casual
inspection, the results of Fig.~\ref{fig:qchem} for $n\ge 5$ are 
decisively improved.   With previous procedures, the present
results become qualitatively like the published, and puzzling, results for K$^+$ with $n \ge 5$.\cite{rogers:ARCC} 
Of course, the similarity of Rb$^+$ and K$^+$ is the natural physical expectation.  The implied computed
hydration free energy based on $\bar n = 4$ is not much changed in quality 
compared to estimates based on the earlier procedures,\cite{rogers:ARCC} as also
noted previously.\cite{svarm08}  For example,
$\mu_{\mathrm{Rb^+}}^{\mathrm{{(ex)}}}$ is only 2~kcal/mol more
positive  using Eq.~\ref{eq:pqca} based on a single cluster geometry and electronic charge distribution
independent of the environment.  The difference here is that
now the overall occupancy distribution  (Fig.~\ref{fig:qchem}) 
is also qualitatively reasonable.

This \emph{single-point} estimate of the cluster free energy was tested
by sampling inner-shell structures from the AIMD simulation record and
evaluating electrostatic contributions to the free energies using a
simple dielectric continuum solvation model.   In that case, the excess
free energies of the sampled clusters and individual water ligands were
estimated using the APBS-version 1.3 software\cite{BAKERNA01A} with the same
parameters described earlier.\cite{svarm08}
%with partial atomic charges required
%for numerical solution of the Poisson equation obtained by the ChelpG
%method\cite{BRENEMAN90A} and atomic radii for hydrogen and oxygen taken from Ref. \cite{STEFANOVIC95A}.
Results for the  sampled clusters were then  combined in the
thermodynamically consistent fashion, \emph{i.e.,} by adding the
inverses of the Boltzmann factors.  That test produced slightly lower
outer-shell electrostatic contributions and $\mu^\mathrm{{(ex)}}_{\mathrm{Rb}^+}$(aq)$ = -70.5$~kcal/mol,
thus confirming the results above.
The conclusion appears to be that isolated $n\ge 5$ clusters are sufficiently
unusual in structure to cause trouble for single-point estimates
of $\mu_{\mathrm{Rb (H_2O)_{n}^+}}^{\mathrm{{(ex)}}}$ based on gas-phase structures.  
Single solvent-adapted structures, or structures sampled from 
liquid-phase simulations, result in a satisfactory improvement.

%\input{R&D.tex}
%\section{Results and discussion} \label{sec:results}

\section{Concluding Discussion} \label{sec:conclude}

%non-split occupancy, experiment, single configs affected by environment vs multiple configs from liquid simulation
%max in Rb-O distribution, <n> at ....,  
%compare with to K+ and contrast with Na+
Using AIMD simulation and primitive quasi-chemical theory,
we find that $n=4$ waters preferentially solvate Rb$^+$ within
a spherical inner shell defined by radius $\lambda=3.2~\AA$.
This boundary extends slightly beyond the first peak of the Rb-O radial
distribution function observed in AIMD simulations and reported in 
experiments ($r$ = 3.0~{\AA}), but lies well within
the putative minimum at $r$ = 3.8~{\AA} occupied by $n$ = 7 waters.  
As customary, we intentionally chose a smaller inner hydration shell for free energy analysis to avoid
split occupancy between first and second hydration shells
observed for the 7th-most distant water from Rb$^+$
in AIMD simulations.  

The hydration free energy for Rb$^+$ predicted by primitive quasi-chemical evaluation agrees
reasonably well with experiment.  Good agreement is achieved when
treating the most probable aqueous coordination complex within
the defined inner hydration shell as a \emph {single} energy-optimized gas-phase species using quantum mechanical
methods and coupling that cluster to an implicit model of the solution.   
For example, coupling  Rb$^+$(H$_2$O)$_4$ to a dielectric continuum model, whether or not the
cluster adapts to that environment, results in a hydration
free energy of $\approx -65$ kcal/mol.    In comparison, coupling many $n = 4$ configurations sampled from
AIMD simulation  to an implicit solvation model results in a similar hydration free energy prediction of -70.5 kcal/mol,
a result within 1 kcal/mol of experiment.

Although the overall hydration free energy remains similar,
a significant improvement does occur in the distribution of inner-shell coordinations predicted when treating solvent-adapted clusters (one or many) compared to prior procedures that treated only
gas-phase clusters.
Comparison of the occupancy distribution from
AIMD with those from a LJ model indicates that AIMD results exhibit
slightly lower coordination numbers with greater variability.

Based on similarity in size (within 0.2~\AA) and identical charge, Rb$^+$ and K$^+$ ions are expected to 
share similar solvation characteristics in water and other environments.   Foremost, hydration 
free energies 
are nearly identical according to experiments, a result also predicted by quasi-chemical analyses.
Further similarities can be highlighted by comparing AIMD simulation and quasi-chemical free energy analysis
of Rb$^+$  with earlier studies of K$^+$.\cite{REMPE04A,VARMA06B,Varma:07bj,svarm08,rogers:ARCC}  
 For example, the first peaks in ion-O radial distibution functions fall in similar locations, 
but slightly closer for the smaller K$^+$ ion ($r$ = 2.8~{\AA}), as anticipated.
In both cases, a weak minimum obscures identification of a coordination shell.
For K$^+$, this minimum ostensibly occurs closer ($r = 3.5~{\AA}$) with occupation by $n$ = 6 waters, 
one less than Rb$^+$.
Similar to Rb$^+$, split occupancy observed in the AIMD record of K$^+$(aq) simulations, 
specifically relating to the 6th-most distant water, motivated
definition of a more restricted inner hydration shell for free energy analysis.

Perhaps unexpectedly, these inner shells are 
defined at nearly identical distances  for  Rb$^+$ and K$^+$ ions ($\lambda_{K^+}=3.1~{\AA}$)
and have similar properties: the first $n$=1-4 waters 
fill-in the first peaks in the ion-oxygen radial distribution functions, 
and $n$=4 waters preferentially solvate both K$^+$ and Rb$^+$ according to 
AIMD simulation and quasi-chemical analysis.    In contrast, $n$=4 waters fill-in the first peak
and preferentially
solvate Na$^+$ within a closer inner-shell distance of $\lambda$ = 2.6~\AA.\cite{svarm08,rogers:ARCC}
Finally,   stable inner-shell hydration structures for both K$^+$ and Rb$^+$ exist in gas phase
that are absent in AIMD simulations of the liquid phase.   For example, the
$n$ = 8 non-split occupancy is a rare composition in AIMD simulation, but forms a
stable skewed cubic structure with 4 waters in a plane above and 4 below  K$^+$\cite{svarm08} 
and Rb$^+$\cite{Ortega:09}
 in the absence of stabilizing interactions with the more distant solvation environment.   
This high coordination contrasts with smaller  
inner-shell coordinations ($n \le 6$) reported for Na$^+$.\cite{svarm08,rogers:ARCC}
Further, these 8-coordinate clusters resemble the crystallographic ligands resolved around K$^+$ and Rb$^+$ ions in the
binding sites of potassium ion channels that conduct K$^+$ and Rb$^+$, but reject
smaller Na$^+$ ions.\cite{Zhou:2001,mackinnon:energy:01}  Following previous studies of selective K$^+$ and 
Na$^+$ binding,\cite{Varma:07bj,svarm08,VARMA08A,varma:jgp} quasi-chemical theory may be useful in 
future work to analyze the subtleties of  Rb$^+$ binding and conduction in potassium channels.

Quasi-chemical theory applied to ion hydration combines statistical
mechanical theory, electronic structure calculations, and molecular
simulation, disciplines which are individually subjects for specialized
professional attention. Because it combines activities which are
themselves non-trivial, quasi-chemical theory is typically viewed with
surprise.  Nevertheless, it provides a
fully-considered framework for analysis of ion hydration. 

It is striking that the three sub-disciplines noted (statistical
mechanical theory, electronic structure calculations, and molecular
simulation) are so distinct. Typical practice in each subdiscipline is
to parameterize the ingredients from the other two in order to eliminate
those complexities. Thus, for example,  sophisticated electronic
structure calculations are done with solution models (dielectric
continuum models) that are not justified on the basis of more basic
observation. Similarly, sophisticated statistical mechanical theory is
typically pursued where molecular-scale realism of the model can be
empirically eliminated, for example by treating parameterized
pair-decomposable models of intermolecular interactions. Simulation
calculations also, of course, adopt extensively parameterized models. 
But they also have the limitation of being non-theoretical, \emph{i.e.,}
not requiring physical insight they most often do not result in any. 
Indeed, simulations can be high-resolution experiments of undetermined 
accuracy for any physical system.

Quasi-chemical theory  is not a \emph{take-it-or-leave-it} model, and
not a series expansion, but a well-defined structure for combining
computational results from distinct sources that treat separately near
and more distant neighbors. As more advanced applications are
encountered --- here with the Rb$^+$(aq), which localizes near-neighbor
water molecules slightly less definitely than some less advanced cases
--- some physical learning and judgement is required. A big step in that
learning has been to catagorize near-neighbors on the basis of 
the neighborship decomposition of the radial distribution as in Fig.~\ref{fig:pgr}. 
This is in contrast to identification of neighbors  on the basis of 
the location of the first  minimum of that radial distribution, which is
often less than compelling.  Another step in that learning has been
to consider more sophisticated procedures for estimation of 
cluster hydration free energies.    Our discussion here has emphasized
further the clear learning point that fluctuations of the 
\emph{composition} of the inner-shell are numerically non-significant
for strongly bound cases such as Rb$^+$ and where a well-informed 
identification of inner-shell ligands has been achieved.

\vspace{0.5cm}
\pagebreak
{\bf Acknowledgment}

Sandia National Laboratories is a multiprogram laboratory managed and
operated by Sandia Corporation, a wholly owned subsidiary of
Lockheed Martin Corporation, for the U.S. Department of Energy's National
Nuclear Security Administration under Contract DE-AC04-94AL8500.
This work was supported by Sandia's LDRD program and by the National
Science Foundation under the NSF EPSCoR Cooperative Agreement No.
EPS-1003897, with additional support from the Louisiana Board of
Regents.

%\bibliography{sabo,RTIL,multistate,bib,reference-supercapacitor_02}

\providecommand*{\mcitethebibliography}{\thebibliography}
\csname @ifundefined\endcsname{endmcitethebibliography}
{\let\endmcitethebibliography\endthebibliography}{}

\footnotesize{
%\bibliography{rsc} %your .bib file
\bibliographystyle{rsc} %the RSC's .bst file
}

\end{document}